\documentclass[aps,showpacs,preprint,superscriptaddress]{revtex4}
\usepackage{graphicx}
\begin{document}
\bibliographystyle{revtex}
\title{Microscopic  study of collective excitations in rotating 
nuclei}
\vspace{0.5cm}
\author{J. Kvasil}
\affiliation{Institute of Particle and Nuclear Physics, Charles
University, V.Hole\v sovi\v ck\'ach 2, CZ-18000 Praha 8, Czech 
Republic}
\author{N. Lo Iudice}
\affiliation{Dipartimento di Scienze Fisiche, Universit\'a di Napoli 
"Federico II" \\and 
Istituto Nazionale di Fisica Nucleare,\\
Monte S Angelo, Via Cinthia I-80126 Napoli, Italy}
\author{R. G.~Nazmitdinov}
\affiliation{Departament de F{\'\i}sica, 
Universitat de les Illes Balears, E-07122 Palma de Mallorca, Spain}
\affiliation{Bogoliubov Laboratory of Theoretical Physics,
Joint Institute for Nuclear Research, 141980 Dubna, Russia}
\author{A. Porrino}
\affiliation{Dipartimento di Scienze Fisiche, Universit\'a di Napoli 
"Federico II" \\and 
Istituto Nazionale di Fisica Nucleare,\\
Monte S Angelo, Via Cinthia I-80126 Napoli, Italy}
\author{F. Knapp}
\affiliation{Institute of Particle and Nuclear Physics, Charles
University, V.Hole\v sovi\v ck\'ach 2, CZ-18000 Praha 8, Czech 
Republic}

\date{\today}
\begin{abstract}
We have carried out a unified microscopic study of  electric monopole,
quadrupole and magnetic dipole excitations in fast rotating nuclei
undergoing backbending, with special attention at the magnetic 
excitations.
We found, among other results, that the strength of the orbital  
magnetic dipole  excitations (scissors mode)
gets enhanced by more than a factor four at high rotational frequency, 
above the backbending region.
We provide a physical explanation for such an enhancement.   
\end{abstract}
\maketitle

\section{Introduction}
Deformation is known to affect deeply  
the collective 
nuclear motion \cite{BoMoII}.  
We mention here the split of the electric 
giant dipole (E1) \cite{Dan58,Oka58}, 
quadrupole (E2) \cite{Kish75} and octupole (E3) \cite{oct92,n92} 
resonances, as well as the coupling between 
quadrupole and monopole (E0)  modes \cite{SuRo77,ZaSpet78,Ab87}. 
It generates also a completely new excitation, 
the scissors mode 
\cite{Lo78,Boh84,Lo00}. 

With the advent of heavy-ion accelerators and of a new generation of detectors,
it was possible to get access to fast rotating nuclei and to observe quite new phenomena induced by the rapid rotation.  
Backbending is a well known spectacular example  \cite{RinSch}.
Systematic theoretical investigations have clarified to a great extent how fast
rotation affects most of the nuclear properties,     
including $\beta-$ and $\gamma-$ modes
\cite{KN,Sh95}, low-lying
octupole excitations and alignment \cite{Rob,Na87,Na96},
pairing vibrations \cite{RMP,AF01}. 
All these studies were carried out in cranked 
random-phase-approximation (CRPA) using separable effective interactions.
The same approach was adopted for extensive studies of the
E1 giant resonance \cite{Sno86,Gaar}.

Less explored is the effect of rotation on other collective 
excitations. To our knowledge, monopole and quadrupole resonances 
were studied only in \cite{Sh84} within the CRPA, 
using the cranked modified harmonic oscillator (HO), 
and in \cite{Ab87} within a phonon plus 
rotor model, using  schematic RPA to generate the phonons.
  
In the present paper, we intend to complete the analysis of Refs. 
\cite{Ab87,Sh84} by including the study of the magnetic 
dipole (M1) excitations, specially the orbital ones known as scissors mode.  
This, in fact, not only is  
intimately linked to deformation and, more in general,
to quadrupole correlations, but, by its own nature, 
is also strictly correlated 
with nuclear rotation. 
Its properties  might therefore change 
considerably in going from slow to fast rotating nuclei.

Our procedure is framed within the CRPA 
and parallels closely the model of Ref. \cite{Na96}. 
We adopted, in fact, a cranked Nilsson model plus quasiparticle RPA and 
made use of doubly stretched 
coordinates. There are, on the other hand, several differences  
concerning mainly  the choice and treatment of the Hamiltonian 
as well as  the method of computing the 
electromagnetic response.  

We applied our method to $^{156}$Dy and $^{158}$Er. 
The evolution of their moment of inertia with the 
rotational frequency was studied within an approach using  the same 
mean field adopted here and found to be consistent with the behavior 
observed experimentally, including the backbending region \cite{KvNa2}. 
This strengthens our confidence on the reliability of our predictions 
on the M1 mode, whose properties, as we shall see, depend critically on
the nuclear moment of inertia.

\section{RPA in the rotating frame}

\subsection{The Hamiltonian}
We started with the Hamiltonian  
\begin{equation}
\label{3}
H_{\Omega} \,=\, H - \hbar \Omega \hat I_1 \,=\,
H_0 - \sum_{\tau=n,p} \lambda_{\tau} N_{\tau} - \hbar \Omega I_1 + V.
\end{equation}
The unperturbed piece is composed of two terms  
\begin{equation}
H_0=\sum_{i} (h_{Nil}(i) + h_{add}(i)). 
\end{equation}
The first is the Nilsson Hamiltonian
\begin{equation}
\label{4}
h_{Nil} =  \frac{p^2}{2m} + V_{HO} + 
v_{ls} {\bf l} \cdot {\bf s} +  v_{ll}({\bf l}^{2} - <{\bf l}^{2}>_N),
\end{equation}
where
\begin{equation}
V_{HO} = \frac{1}{2} m (\omega_1^2 x_{1}^2 + \omega_2^2 x_{2}^2 + 
\omega_3^2 x_{3}^2)
\label{TriHO} 
\end{equation}
is a triaxial HO, whose frequencies satisfy the volume 
conserving condition $\omega_{1} \omega_{2} \omega_{3} = \omega_0^3$.
The second comes from  restoring  the local 
Galilean invariance broken in the rotating coordinate 
systems and has the form \cite{Na96}   
\begin{eqnarray}
\label{4a}
h_{add} = - \frac{\Omega}{\sqrt{\omega_{2} \omega_{3}}}   
\left\{v_{ll} \left[2 m \omega_0 {\bf r}{^\prime}^2 - 
\hbar \left( N_{(osc)} + 
\frac{3}{2}\right)\right] l^\prime_{1} 
+ v_{ls} m \omega_0 \left[ \bf r^{\prime^2} s_{1} -
x_{1}^\prime (\bf r^\prime \cdot \bf s) \right] \right\}, 
\end{eqnarray}
where $x^\prime_i = (\omega_i/\omega_0)^{1/2} x_i$ 
are single-stretched coordinates.

The two-body potential has the following structure 
\begin{equation}
V = V_{PP} + V_{QQ} + V_{MM} + W_{\sigma \sigma}.
\end{equation}
$V_{PP}$ is a monopole pairing interaction
\begin{equation}
V_{PP} = - \sum_{\tau=p,n} G_{\tau} P_{\tau}^\dagger P_{\tau},
\end{equation}
where $P_{\tau}^{\dagger} = \sum_{k} a_k^\dagger a_{\bar k}^{\dagger}$ is the 
usual pairing operator.
$V_{QQ}$ and $V_{MM}$ are, respectively, 
separable quadrupole-quadrupole and monopole-monopole interactions
\begin{eqnarray}
V_{QQ} &=& - \frac{1}{2} \sum_{T=0,1} \kappa (T) 
\sum_{r=\pm} \sum_{\mu=0,1,2} (\tilde Q_\mu [^{T}_{r}])^2,\nonumber\\
V_{MM} &=& - \frac{1}{2} \sum_{T=0,1} \kappa (T) (\tilde M 
[^{T}_{r=+}])^2.
\end{eqnarray}
$V_{\sigma \sigma}$ is a spin-spin interaction 
\begin{equation}
V_{\sigma \sigma} = - \frac{1}{2}  
\sum_{T=0,1} \kappa_{\sigma} (T)
\sum_{r=\pm} \sum_{\mu=0,1} (s_\mu [^{T}_{r}])^2.
\end{equation}
Because of its repulsive character, this interaction pushes the spin excitations at higher energy, thereby generating two well separated regions, one below 4 MeV, composed of  mainly orbital excitations and another, in the range 4 MeV $\div$ 12 MeV, characterized by spin excitations \cite{DeHe89,DeHe91}.
    
All the one-body fields have good isospin $T$ and signature $r$.
Multipole and spin-multipole fields of good 
signature are defined in Ref. \cite{Kv98}.
The tilde indicates that monopole and quadrupole fields are 
expressed in terms of doubly stretched coordinates
$x_i^"=(\omega_i/\omega_0)\,x_i$ \cite{Kish75,SaKi89}. 
In this new form, for a pure 
HO Hamiltonian, 
the quadrupole fields fulfill the stability conditions
\begin{equation}
<\tilde Q_\mu> = 0, \qquad \mu=0,1,2
\label{new}
\end{equation}
if nuclear self-consistency  
\begin{equation}
\omega_{1}^2 <x_{1}^2> = \omega_{2}^2 <x_{2}^2> = 
\omega_{3}^2 <x_{3}^2> 
\end{equation}
is satisfied in addition to the volume conserving condition. 
In virtue of the constrains (\ref{new}),  
the interaction will not distort further the deformed HO potential, 
if the latter is generated  as a  Hartree field.  This is achieved by starting 
with an isotropic HO potential of frequency $\omega_0$ and, then, 
by generating 
the deformed part of the potential from the (unstretched) 
quadrupole-quadrupole interaction. The outcome of this procedure is  
\begin{equation}
V_{HO} = \frac{m\omega_0^2 r^2}{2} - m\omega_0^2 \beta \,cos\gamma \, 
Q_{0}[^{\,0}_{+}] - m\omega_0^2 \beta \,sin\gamma 
Q_{2}[^{\,0}_{+}]
\end{equation}
where
\begin{eqnarray}
m\omega_0^2 \beta \,cos\gamma &=&   
 \kappa[0] <Q_{0} [^{\,0}_{+}]>\nonumber\\
m\omega_0^2 \beta \,sin\gamma &= &\kappa[0] <Q_{2} [^{\,0}_{+}]>
\label{Hartree}
\end{eqnarray}
The triaxial form given by Eq. (\ref{TriHO}) follows from
defining
\begin{equation}
\omega_i = 
\omega_0 \exp  [-\frac{2}{3} \delta \cos{(\gamma - i \frac{2 
\pi}{3})}],
\,\,\,\,\,\,i=1,2,3,
\end{equation}
where the new deformation parameter is defined by 
$\beta = \sqrt{16 \pi/45} \, \,\delta$.
The Hartree conditions have the form given by Eq.(\ref{Hartree})  
only for an HO plus a
separable quadrupole-quadrupole potential. They  change 
once pairing is included \cite{Lo96}. 
Moreover, they fail to give   
the minimum of the mean field energy of the rotating system
in superdeformed  nuclei \cite{A03}. 
For all these reasons, we have allowed for small deviations from 
Eqs. (\ref{Hartree}) and enforced, instead, 
 the stability conditions
(\ref{new}). These, in fact, hold also in the presence of pairing   
\cite{KvNa2} and guarantee 
the separation of the pure rotation from the intrinsic 
vibrational modes in the limit of rotating harmonic oscillator \cite{Na02}.

\subsection{Quasiparticle RPA in rotating systems}

We expressed first the Hamiltonian in terms of quasiparticle
creation ($\alpha^\dagger_i$) and 
annihilation ($\alpha_i$) operators by carrying out at 
each rotational frequency $\Omega$ a 
generalized Bogoliubov transformation. 
We then faced the RPA problem 
and wrote the RPA equations in the form \cite{KN,Kv98}
\begin{eqnarray}
\label{10}
[H_{\Omega}, P_{\nu}] \,=\, i\,\hbar \omega_{\nu}^2\,X_{\nu}, && \quad
[H_{\Omega}, X_{\nu}] \,=\, -i\,\hbar \,P_{\nu}, \qquad
[X_{\nu}, P_{\nu'}]\,=\,i \hbar \delta_{\nu\,\nu'},
\end{eqnarray}
where $X_{\nu}$, $P_{\nu}$ are, respectively, 
the collective coordinates 
and their conjugate momenta.  
The solution of the above equations yields the RPA 
eigenvalues $\hbar \omega_{\nu}$
and eigenfunctions   
\begin{eqnarray}
|\nu> \,= \,O_{\nu}^\dagger |RPA> &=& 
\frac{1}{\sqrt{2}}\,\Bigl(\,\sqrt{\frac{\omega_{\nu}}{\hbar}} 
\,X_{\nu} \,-\,
\frac{i}{\sqrt{\hbar \omega_{\nu}}}\,\hat{P}_{\nu} \,\Bigr) 
|RPA>\nonumber\\
&=& \sum_{ij} \Bigl(\psi_{ij}^\nu b^\dagger_{ij} - \Phi^\nu_{ij} 
b_{ij} \Bigr) |RPA>,
\end{eqnarray} 
where 
$b^\dagger_{ij} = \alpha_i^\dagger \alpha_j^\dagger \,\,\, (b_{ij} = 
\alpha_i \alpha_j)$ 
creates (destroys)  a pair of quasiparticles out of the RPA vacuum  
$\mid RPA \rangle$ .
Since the Hamiltonian can be decomposed into the sum of a positive and
a negative signature piece
\begin{equation}
H_{\Omega} = H_{\Omega} (r=+) + H_{\Omega}(r=-),
\end{equation}
we solved the eigenvalue equations (\ref{10})  
for $H_{\Omega} (+)$ and $H_{\Omega} (-)$ separately.

The symmetry properties of the cranking Hamiltonian yield  
\begin{eqnarray}
\label{spurious1}
&&[\, H \,,\,N_{\tau=n,p}\,]_{RPA}\, =  \,0,  
\,\,\,\,[H\,,\,I_1\,]_{RPA}=0   
\\ 
&&[\, H \,,\,I_2 \,]_{RPA}\, = i \hbar \Omega I_3,        
\,\,\,\,[\,H\,,\,I_3\,]_{RPA}\,=\,-i \hbar \Omega I_2.
\label{spurious2}
\end{eqnarray}
The last two equations can be combined so as to obtain
\begin{equation}
[H_\Omega(-),\Gamma^{\dagger}]=\Omega \Gamma^{\dagger}
\label{rot}
\end{equation}
where  $\Gamma^{\dagger}=(I_2 + i I_3)/\sqrt{2 \langle I_1 \rangle}$
and $\Gamma = (\Gamma^{\dagger})^{\dagger}=
(I_2 - i I_3)/\sqrt{2 \langle I_1 \rangle}$
fulfill the  commutation relation
\begin{equation}
[\Gamma, \Gamma^{\dagger}]=1. 
\end{equation}
According to Eqs. (\ref{spurious1}), 
we have two Goldstone modes, one
associated to the violation of the particle number 
operator, the other is a positive signature 
zero frequency rotational solution associate to
the breaking of spherical symmetry. 
Eq. (\ref{rot}), on the other hand, yields  
a negative signature  redundant solution 
of energy $\omega_{\lambda}=\Omega$, which describes  
a collective rotational mode arising from the symmetries broken  
by the external 
rotational field (the cranking term).  

Eqs. (\ref{spurious1}) and (\ref{rot}) guarantee 
the separation of the spurious or redundant solutions 
from the intrinsic ones. They would be automatically 
satisfied if the single-particle basis were generated through a 
self-consistent Hartree-Bogoliubov (HB) calculation. 
As we shall show, they are fulfilled with good accuracy also 
in our, not fully self-consistent, HB treatment.

The strength function for an electric ($X=E$) or magnetic ($X=M$) 
transition of multipolarity $\lambda$ from 
a state of the yrast line with angular momentum $I$ is 
\begin{equation}
\label{2}
S_{X\lambda}(E) \,=\, \sum_{\nu \, I^\prime} B(X\lambda,\,I\, 
\rightarrow
\,I^\prime,\,\nu)\,\delta(E-\hbar \omega_{\nu}),
\end{equation}
where  $\nu$  labels all the  excited states with a given $I^\prime$.
In order to compute the reduced strength $B(X\lambda,\,I\, \rightarrow
\,I^\prime,\,\nu)$ we should be able to expand the intrinsic RPA 
state  into components with good 
$K$ quantum numbers, which is practically impossible in the cranking 
approach. 
We have, therefore, computed the strength in the limits of  zero and 
high  frequencies.
For non rotating axially symmetric nuclei, whose 
initial state is usually the $I=0, K^{\pi}\,=\,0^{+}_{gr}$ ground 
state, the strength function is given by
\begin{equation}
\label{1}
S_{X\lambda}(E) \,=\, \sum_{\nu K} B(X\lambda,\,0^{+}_{gr} 
\rightarrow\,K\nu)\,\delta(E-\hbar \omega_{\nu}),
\end{equation}
where  
\begin{equation}
\label{11}
B(X\lambda,\,0_{gr}^+ \rightarrow K_{\nu})\,=\, 
|\,<RPA|\,[O_{K\nu},{\cal M}(X\lambda \mu_{3}=K)]\,|RPA>\,|^2.
\end{equation}
For fast rotating nuclei, we assumed a complete alignment 
of the angular momentum
along the rotational $x_1$-axis,  so that  ($I^\prime = I + \Delta I$)
\begin{equation}
\label{2a}
S_{X\lambda,\,\Omega(I)}(E) \,=\, 
\sum_{\nu\,\Delta I} B(X\lambda,\,I\,yrast \rightarrow
\,I+\Delta I,\,\nu r)\,\delta(E-\hbar \omega_{\nu})
\end{equation}
where ($\Delta I \, =\, 0,\pm 1,....,\pm \lambda$)  
\begin{eqnarray}
\label{12}
B&&\!\!\!\!\!(X\lambda,I\rightarrow I\!+\!\Delta I, \,\nu r) =
\nonumber\\
&&
=\,|\,(I\,I\,\lambda\,\Delta I\,|\,I\!+\!\Delta I\,I\!+\!\Delta 
I\,)\,\,
_{\Omega}\!\!<RPA|\,[O_{\nu r}\,,\, {\cal M}(X\lambda \mu_{1}=\Delta 
I)]\,
|RPA>_{\Omega}\,|^2,
\nonumber\\
\end{eqnarray}
having denoted by $|RPA \rangle_{\Omega}$ the RPA vacuum (yrast state) at 
the rotational frequency $\Omega$. 
The multipole operator in the rotating frame  
was obtained from the corresponding one in the laboratory    
according to the prescription \cite{Ma76} 
\begin{equation}
{\cal M}(X\lambda \mu_{1}) = 
\sum_{\mu_{3}} \,\,{\cal D}^\lambda_{\mu_{1} \mu_{3}} 
(0,\frac{\pi}{2}, 
0) \,\,{\cal M} (X \lambda \mu_{3}).
\end{equation}
Having used an interaction composed of a sum of 
separable pieces, we did not need to 
determine explicitly the RPA eigenvalues and 
eigenfunctions in order to calculate the strength function.  
Following the method of Ref. \cite{KN,Kv98}, 
we had just to replace the $\delta$ distribution with 
a Lorentz weight  and, upon the use of the Cauchy theorem, we obtained  
for $S_{X\lambda}(E)$ and
$S_{X\lambda,\,\Omega(I)}(E)$ expressions involving only 
two-quasiparticle matrix
elements of one-body multipole operators.

The n-th moments were obtained simply as   
\begin{equation}
\label{16}
m_n(X \lambda) \,=\, \int_0^{\infty} \, E^n \, S_{X\lambda}(E) \,dE.
\end{equation}
The $m_0(X\lambda)$  and $m_1(X\lambda)$ moments 
give, respectively, the energy unweighted and weighted summed strengths.

\section{Numerical calculations and  results}

\subsection{Determination of the parameters}
We took the parameters of the rotating Nilsson potential
from Ref.  \cite{Ja90}. They were determined from a systematic 
analysis of the experimental single-particle levels of 
deformed nuclei of rare earth and actinide regions. 
We included all shells 
up to $N=8$ and accounted for  the $\Delta N=2$ mixing.

In principle, the pairing gap should have been determined 
self-consistently at each rotational frequency. 
In order to avoid unwanted singularities for certain 
values of $\Omega$, we followed  
the phenomenological prescription \cite{Wyss}
\begin{equation}
\label{17}
\Delta_{\tau}(\Omega) \,=\,
\left \{
\begin{array}{l}
\Delta_{\tau}(0)\,[1-\frac{1}{2} (\frac{\Omega}{\Omega_c})^2 \,] 
\qquad 
\,\,\,\,\,\,\Omega < \Omega_c \\
\Delta_{\tau}(0)\,
\frac{1}{2} (\frac{\Omega_{c}}{\Omega})^2 \qquad \qquad \, \,  
\,\,\,\,\,\,\,\,\Omega > \Omega_c \\
\end{array},
\right.
\end{equation}
where $\Omega_c$ is  the critical rotational frequency 
of the first band crossing.  
We got, for both neutrons and protons, 
$\Omega_c=0.32$ MeV for $^{156}Dy$ and $\Omega_c=0.33$ MeV
for $^{158}Er$.
We extracted the values of the pairing gaps at zero 
rotational frequency  from the odd-even mass difference 
obtaining $\Delta_N(0)=0.857$ MeV, $\Delta_P(0)=0.879$ MeV 
for $^{156}Dy$ and $\Delta_N(0)=0.874$ MeV, $\Delta_P(0)=0.884$ MeV 
for $^{158}Er$.
\begin{figure}[ht]
\includegraphics[height = 0.25\textheight]{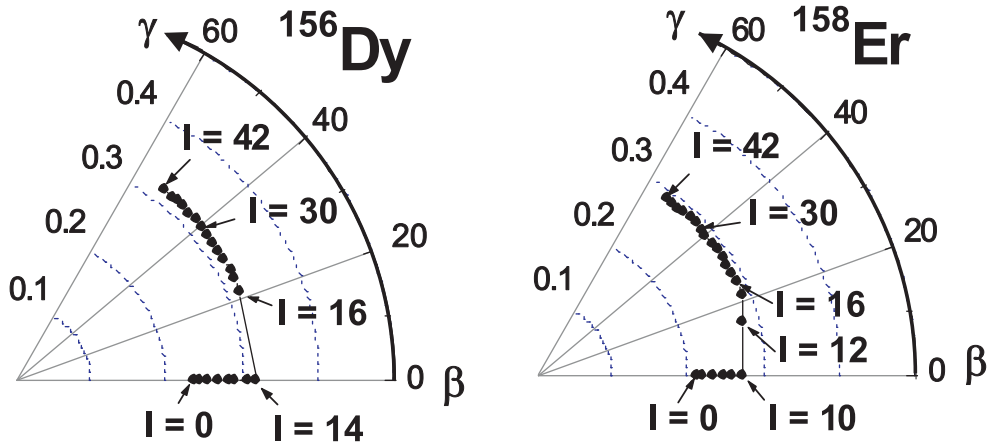}
\caption{(Color online) Yrast line deformations in $\beta$-$\gamma$ plane
as a function of the angular momentum.}
\label{fig1}
\end{figure}
We used as input for our HB calculations  
the deformation parameters obtained from the empirical moments 
of inertia  at each $\Omega$ \cite{Sa99}.
As shown in Fig.\ref{fig1} and discussed elsewhere \cite{KvNa2}, 
triaxiality sets in  at the frequency which triggers  backbending 
as result of the vanishing of the gamma excitations of  positive signature in the rotating frame. 

The parameters so determined yield results  
in better agreement with experiments compared to  
the predictions of  
Ref. \cite{Frau} for $N \sim 90$, where  a fixed phenomenological
inertial parameter was used for all $\Omega$. 
Moreover, our equilibrium deformations are short from being 
the self-consistent solutions of the HB equations. Indeed,  
any deviation from the equilibrium values of the deformation parameters 
$\beta$ and $\gamma$ resulted into a higher
HB energy. Dealing with transitional nuclei, however, 
the minimum becomes very shallow 
as the rotational frequency increases.
In fact, the energy minima for 
the collective rotations around 
the $x_{1}$ rotational axis and for the non collective ones around
the $x_3$ symmetry axis   
are almost degenerate near the crossing point  
of the ground with  the gamma band.  The energy difference is 
about 15 keV near the critical rotational frequency 
where the backbending occurs. 
At the bifurcation point, the competition between
the collective and non-collective rotations breaks the
axial symmetry and  yields non axial shapes. 
On the other hand, the doubly stretched quadrupole moments are approximately
zero for all values of the equilibrium deformation parameters, 
consistently with the stability conditions (\ref{new}).
A small deviation from the equilibrium deformation yields
a strong deviation of these moments from
zero. We infer from the just discussed tests  
that our solutions are close to the 
self-consistent  HB ones. 

As for the strength $\kappa$ of the monopole and 
quadrupole interactions,  we started with 
adopting the standard HO formulas ($\lambda = 2$) \cite{SaKi89}
\begin{equation}
\kappa_{\lambda}[0]=
\frac{4\pi}{2 \lambda +1} \frac{m\omega_0^2}{A<r^{2\lambda -2}>}, 
\,\,\,\,\,\,\,\kappa_{\lambda}[1]=-\frac{\pi V_1}{A<r^{2\lambda}>}\,.
\label{HOkappa}
\end{equation}
The isoscalar  strength, for instance, follows
from enforcing the Hartree self-consistent conditions. 
We then changed slightly the  strengths at 
each rotational frequency, while keeping constant the 
$\kappa[1]/\kappa[0]$ ratio, so as to fulfill 
the RPA equations (\ref{spurious1})-(\ref{rot}) for the spurious or redundant 
modes. The constants so determined differ from the 
HO  ones by 5-10\% at most. For the spin,  we used the 
generally accepted strengths \cite{Ca76} 
$$
\kappa_{\sigma}[0]\,=\,\kappa_{\sigma}[1]\,=\,- 28 \,\frac{4\pi}{A} 
\, MeV
$$ 
for all rotational frequencies. 
Finally, we used bare charges to compute the $E0$ and $E2$ 
strengths and a quenching factor $g_s = 0.7$ for the 
gyromagnetic ratios to compute the $M1$ strengths.  
 
With the above parameters, it was possible not only 
to separate the spurious and rotational solutions 
from the intrinsic modes, 
but also to reproduce the experimental dependence  
of the lowest $\beta$ and $\gamma$ bands on
$\Omega$ and, 
in particular, to observe the 
crossing of the $\gamma$ with the ground band 
in correspondence of the onset 
of triaxiality \cite{KvNa2}.

\subsection{Evolution of transition strengths with rotation}

\begin{figure}[ht]
\includegraphics[height = 0.45\textheight]{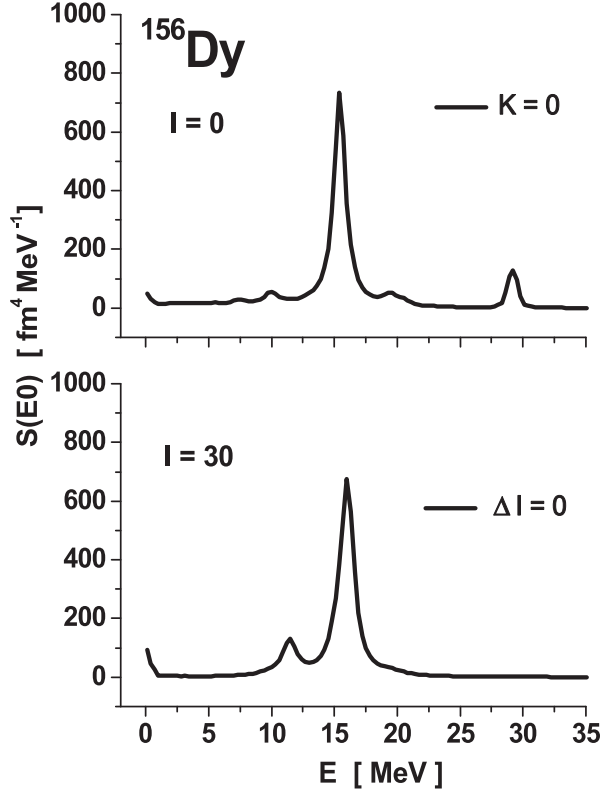}
\caption{E0 strength function  
at zero and high rotational frequencies in $^{156}Dy$.}
\label{fig2}
\end{figure}
We show only the results  of  $^{156}Dy$, since the ones concerning
$^{158}$Er are very similar. As shown in Fig. \ref{fig2},  
the $E0$ response remains unchanged in its dominant isoscalar 
peak.
The effects of fast rotation get manifested through
the suppression of the high energy isovector peak, small in any case,
and the appearance of a peak at $\sim 11 \div 12$ MeV, in correspondence 
of the $K=0$ branch of the quadrupole resonance. This result indicates that  
the coupling between monopole and $K=0$ 
quadrupole modes gets stronger with increasing rotational frequency.
Indeed, as triaxiality sets in at high frequencies, anisotropy 
increases, thereby enhancing the mixing between the two channels.
 
\begin{figure}[hb]
\includegraphics[height = 0.45\textheight]{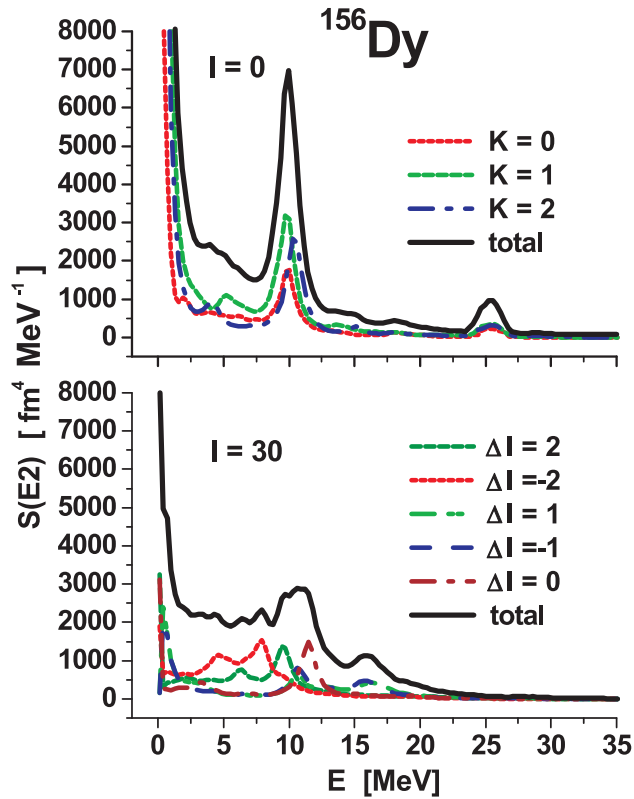}
\caption{(Color online) E2 strength function  
at zero and high rotational frequencies in $^{156}Dy$.}
\label{fig3}
\end{figure}
Rotation has some appreciable effects on the quadrupole 
transitions. 
It broadens considerably the isoscalar quadrupole giant  resonance
because of the  increasing splitting of the different $\Delta I$ 
peaks with increasing $\Omega$.
It washes out  the isovector E2 resonance for the same reason. 
The low-lying peaks shown in Fig. \ref{fig3} are related
to $\gamma$, $\beta$ excitations and collective rotational modes 
described by Eq.(\ref{rot}).
Since these low-lying excitations have been discussed 
elsewhere \cite{KvNa2}, we will confine our study to the excitations 
at higher energy. 
 
The moment $m_1(E2)$ exhausts more than 98\% of the oscillator 
E2 energy-weighted sum rule (EWSR)  
$$
m_1(E2) = \frac{15}{2 \pi} \frac{\hbar^2}{2m} \,A\,R_0^2
$$ 
for all values of $\Omega$. The same result holds for the $E0$ mode.
Thus, the $E0$ and $E2$ EWSR are not affected
by rotation. 

\begin{figure}[ht]
\includegraphics[height = 0.75\textheight]{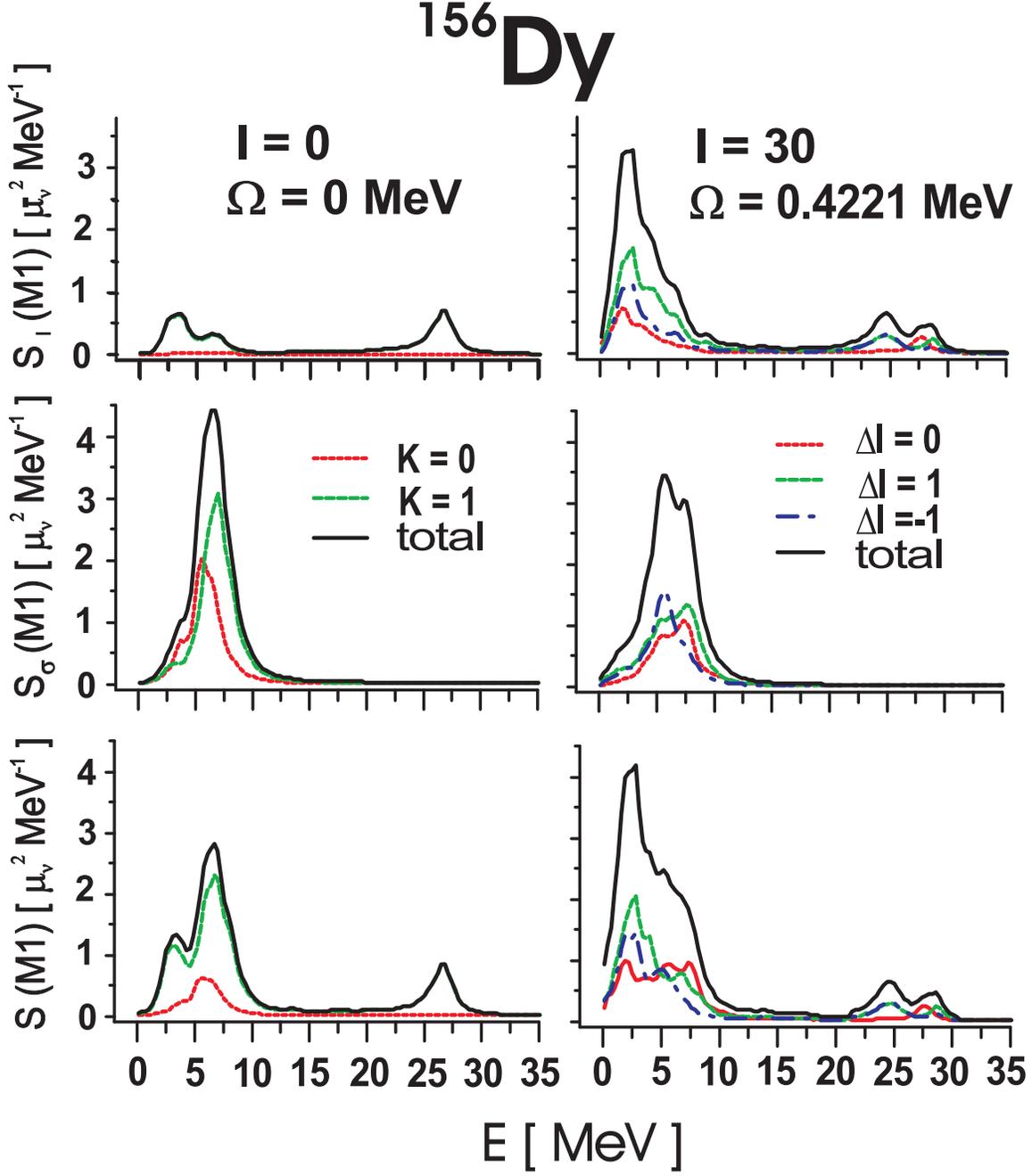}
\caption{(Color online) Orbital, spin, and total M1 strength functions   
at zero (left-hand panels) and high rotational frequencies 
(right-hand panels) in $^{156}Dy$ } 
\label{fig4}
\end{figure}

The strength of the magnetic dipole transitions, at zero 
rotational frequency, is concentrated in three distinct regions,
consistently with the theoretical expectations and the experimental
findings \cite{Lo00}.
We observe a low-energy interval ranging from 2 to 4 Mev, 
characterized by orbital 
excitations (scissors mode \cite{Lo78,Boh84}), a high-energy one 
around 24 MeV  
also arising from magnetic dipole orbital transitions 
(high energy scissors mode \cite{LoRi89}), and
a third intermediate region ranging from 4 to 12 MeV  due to spin 
excitations \cite{Rich93}.

As shown in Fig. \ref{fig4}, the distribution of the strength 
changes considerably as $\Omega$ increases, 
to the point that the dominant peak shifts from 7-8 MeV down to 3 
MeV. Only at high energy, the changes are appreciable but not dramatic.  
In this region, in fact, the M1 strength gets more spread and  
increases  slightly in magnitude (Table \ref{tab1}). 

\begin{figure}[ht]
\includegraphics[height = 0.75\textheight]{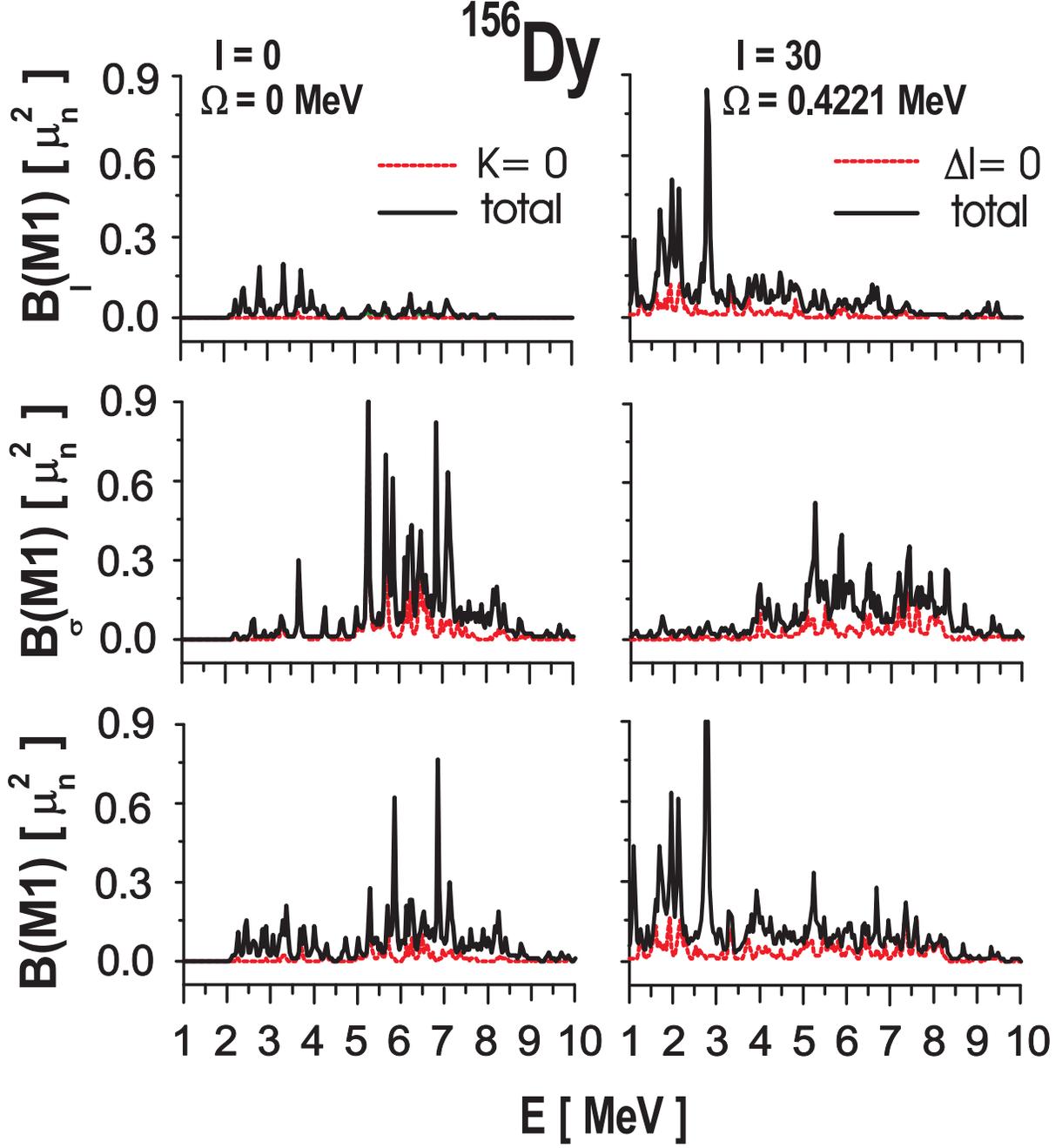}
\caption{(Color online) Orbital, spin, and 
total M1 reduced strength  distributions 
at zero (left-hand panels) and high rotational frequencies (right-hand  panels) in 
$^{156}Dy$.}
\label{fig5}
\end{figure}    
In order to understand what is going on, we analyzed separately 
the contribution of the orbital and spin excitations up to 10 MeV. 
As shown in Fig. \ref{fig5}, the rotation broadens 
the spin strength at the expenses of the main peaks which get severely reduced. The fragmentation keeps the spin transitions confined mainly within  
the range $4\div 12$ MeV (Table \ref{tab1}).  

\begin{figure}[ht]
\includegraphics[height = 0.45\textheight]{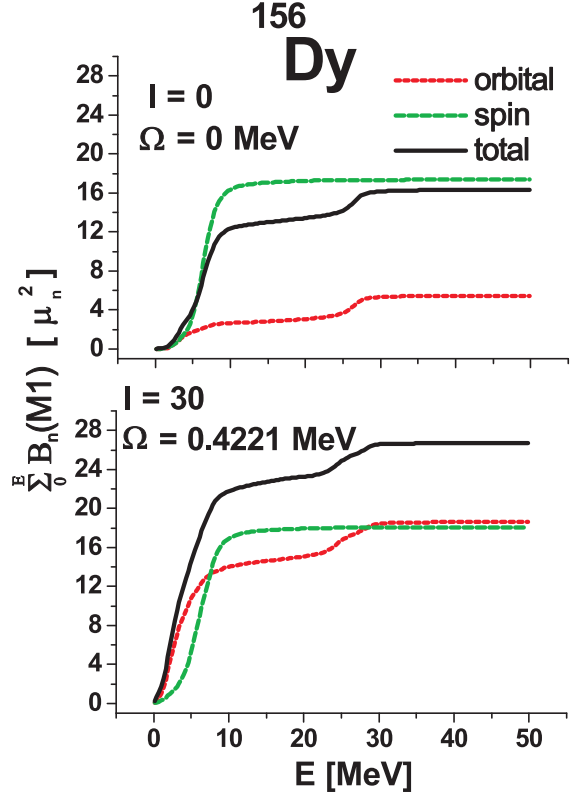}
\caption{(Color online) Running sum of the orbital (dashed line), spin (dashed-dot line) 
and total (solid line) M1 strength in 
$^{156}Dy$.}
\label{fig6}
\end{figure}

%thereby inducing a quenching of 
%the overall strength. 
\begin{table} 
\caption{\label{tab1}Orbital, spin and total M1 strengths integrated over different energy ranges at zero and high ($\Omega = 0.4221$ MeV) rotational frequencies.} 
%\begin{ruletabular}
%{|l|c|c|c|c|c|c|}
\begin{tabular}{|l|c|c|c|c|c|c|}
\hline & \multicolumn{2}{|c|}{ $\,$ 1MeV$ <$E$<$ 4MeV $\,$}&
\multicolumn{2}{|c|}{ $\,$ 4MeV$<$E$<$ 12MeV} &
\multicolumn{2}{|c|}{E $>$ 12MeV} \\
\cline{2-7}
& $\quad I=0\quad$   & $I=30$  
& $\quad I=0\quad$   & $I=30$ 
& $\quad I=0\quad$   & $\,\,I=30\,\,$ \\ 
\hline
$\sum B_l(M1) \,\,[\mu_{n}^{2}]$ & 1.36 & 8.87 & 1.43& 5.39 & 2.75 & 4.35 \\ 
$\sum B_\sigma(M1)\,\,[\mu_{n}^{2}]$ & 1.92 & 3.08 & 14.97 & 14.47 & 0.48 & 0.68 \\ 
$\sum B(M1)\,\,[\mu_{n}^{2}]$ & 2.95 & 12.21 & 9.82 & 10.06 & 3.57 & 4.58 \\
\hline
\end{tabular} 
%\end{ruletabular}
\end{table}
The low-lying orbital strength gets  larger and larger with 
increasing $\Omega$.   
At $\Omega=0$,  the orbital peaks are small compared to the spin 
transitions 
which dominate the M1 spectrum. At $I=30 \hbar$, instead, the  orbital spectrum
not only covers a wider energy range but gets magnified, specially 
in the low-energy sector, where quite high peaks appear.  The low-lying
orbital strength gets enhanced  by 
more than a factor six because of the rotation (Table \ref{tab1}).  
One may also observe that the $\Delta I = 0$ transitions, absent at 
zero frequency ($\Delta I = K = 0$), give a small but non zero 
contribution which increases with $\Omega$. This is due to a new  
branch of the scissors mode which arises with the onset 
of triaxiality \cite{Pa85,Lo85}.  Indeed, in going
from axial to triaxial nuclei, the mode splits into two branches of energy
and M1 strength
\begin{eqnarray}
E_{i} &=& \cos{\gamma} \left[1- \left(-1\right)^{i} 
\frac{1}{\sqrt{3}}\tan{\gamma}\right]E_{sc},\nonumber\\
B_{i}(M1)&=& \frac{1}{2}\cos{\gamma} \left[1- \left(-1\right)^{i} 
\frac{1}{\sqrt{3}}\tan{\gamma}\right] B_{sc}(M1),\,\,\,\,\,\,\,i=1,2
\label{B12}
\end{eqnarray}
where $E_{sc}$ and $ B_{sc}(M1)$ are the energy and strength 
in the axial case. These two branches describe the rotational oscillations around the $x_1$ and $x_2$ axes. A new $K=0$ branch also arises due to the rotational oscillation around the $x_3$ axis. Its energy and 
strength are given by
\begin{eqnarray}
E_{3} &=&  \frac{2}{\sqrt{3}} \sin{\gamma} E_{sc},\nonumber\\
B_{3}(M1)\uparrow &= & 
\frac{2}{\sqrt{3}}\sin{\gamma} B_{sc}(M1).
\label{B3}
\end{eqnarray}
The increasing role of the orbital motion with increasing rotational 
frequency can be also inferred from the plot of the running sums shown in 
Fig. \ref{fig6}.
The orbital strength, small at zero frequency in the whole energy 
range, at high frequencies  becomes by far larger than the 
spin strength in the low-energy 
sector.  
 
\begin{figure}[ht]
\includegraphics[height = 0.5\textheight]{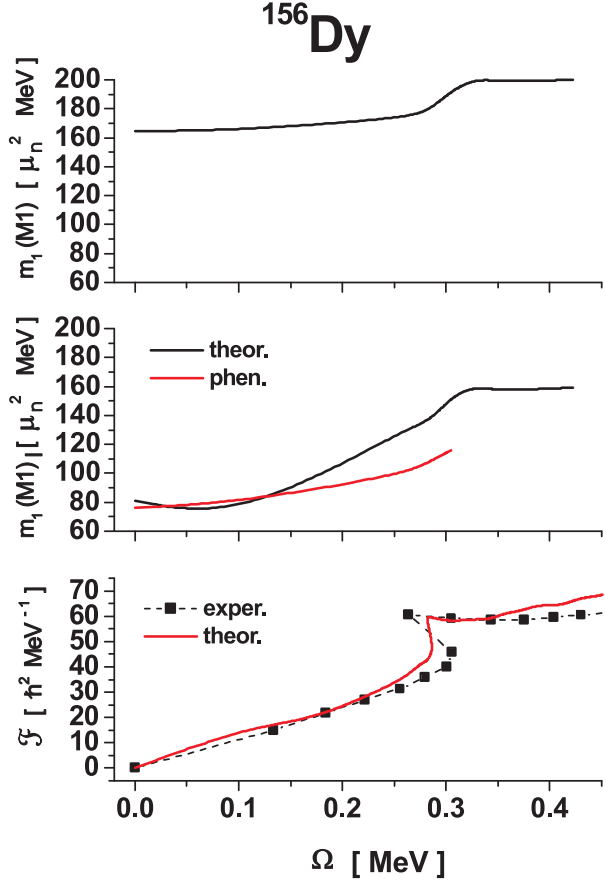}
\caption{(Color online) 
Total (top panel), orbital (middle panel) $m_{1}(M1)$ moments 
and the kinematical moment of inertia (bottom panel) versus $\Omega$  
in $^{156}Dy$. The dashed-dot line displays the M1 EWSR at zero 
frequency computed from Eq. (\ref{M1EW}) (taken from Ref. \cite{Lo98}).}
\label{fig7}
\end{figure}

We can identify one of the mechanisms responsible for such a large 
enhancement by comparing  (Fig. \ref{fig7})
the $\Omega$ dependence of the orbital and total  $m_1(M1)$ moments with  the corresponding evolution of the kinematical moment of inertia $\Im = I/\Omega$, computed using the cranking method of Ref. \cite{A03}. 
The strikingly similar behavior of the orbital $m_1(M1)$  
and the moment of inertia shows that the two quantities 
are closely correlated at all rotational frequencies. 
Indeed, at zero frequency, one has the M1 EWSR \cite{ZaZhe91,Lo98}
\begin{equation}
m_{1}^{(sc)}(M1) = \sum_{n} E_{n} B_{n}^{(sc)}(M1) \simeq 
\frac{9}{16 \pi} \left(\kappa(0) - \kappa(1) \right) \langle Q(0) 
\rangle^{2},\label{M1EW}
\end{equation}
where $Q(0) = Q_{p} + Q_{n}$ is the isoscalar quadrupole field. 
Using the HO formulas (\ref{HOkappa}) for the coupling constants 
and the standard expression for the quadrupole moment \cite{BoMoII}, 
we get for the right-hand side
\begin{equation}
m_{1}^{(sc)}(M1)  \,=\, \frac{3}{8 \pi} (1 - b) \Im_{rig}  
\omega_0^2 \delta^{2},
\label{M1EWSR}
\end{equation}
where $b= \kappa(1)/\kappa(0)$ and 
\begin{equation}
\Im_{rig} = \frac{2}{3} m A <r^{2}>,
\end{equation}
which shows explicitly the close link between the orbital $M1$ EWSR
and the moment of inertia, at zero rotational frequency. 
We get a deeper insight by inspecting more closely the energy unweighted
and weighted sums.  
For both low and high energy modes, the M1 summed strength
has the general form \cite{Lo00}
\begin{equation}
m_{0}^{(\pm)} (M1) =
\sum_{n_{\pm}} B_{n_{\pm}}^{(\pm)} (M1) \simeq \frac{3}{16 \pi} 
\Im^{(\pm)}_{sc} {\bar E}^{(\pm)}, 
\end{equation}
having denoted by  ${\bar E}^{(\pm)}$ and $ \Im^{(\pm)}_{sc}$ the 
energy centroids and the
mass parameters of the high-lying $(+)$ and low-lying $(-)$ scissors 
modes.
At high energy, protons and neutrons behave as normal 
irrotational fluids, so that energy and mass parameter are given by
\begin{eqnarray}
{\bar E}^{(+)}  \propto 2 \omega_{0},\,\,\,\,\,\,\,\,\,\,
\Im_{sc}^{(+)} = \Im_{irr} = \Im_{rig}
\delta^{2},
\end{eqnarray}
yielding  
\begin{eqnarray}
m_{0}^{(+)} (M1) &=&\sum_{n_{+}} B_{n_{+}}^{(+)}(M1)  \propto  \Im_{rig} 
\delta^{2}\nonumber\\
m_{1}^{(+)} (M1)  &\simeq& {\bar E}^{(+)} 
\sum_{n_{+}} B_{n_{+}}^{(+)}  \propto  \Im_{rig} \delta^{2}.
\end{eqnarray}
Thus, both M1 weighted and unweighted summed strengths are quadratic in the deformation parameter. This result remains substantially unchanged in fast rotating nuclei.   

For the low energy mode, instead, we must distinguish between zero and large
rotational frequencies. At zero frequency, protons and neutrons
behave as superfluids, so that \cite{LoRi93,Piet98}  
\begin{eqnarray}
{\bar E}^{(-)} \propto 2 E_{qp} \simeq 2  \Delta,\,\,\,\,\,\,\,\,\,
\Im_{sc}^{(-)} = \Im_{sf} \propto \Im_{rig}
\delta^{2},
\end{eqnarray}
where $ E_{qp}$ denotes the quasiparticle energy and $ \Im_{sf}$ the 
superfluid moment of inertia. We then have
\begin{eqnarray}
m_{0}^{(-)} (M1) &= &\sum_{n_{-}} B_{n_{-}}^{(-)} (M1) \propto  \Im_{rig} 
\delta^{2}\nonumber\\
m_{1}^{(-)} (M1)  &\simeq& {\bar E}^{(-)} 
\sum_{n_{-}} B_{n_{-}}^{(-)} (M1) \propto  \Im_{rig} 
\delta^{2}.\label{mlowom}
\end{eqnarray}
These relations show that  $m_{1} (M1)$ is  consistent with the EWSR
(\ref{M1EW}) and, quite remarkably, the summed strength $m_{0} (M1)$ follows  the quadratic deformation law found experimentally \cite{Zi90,Ma93}.

At high rotational frequency, instead, the pairing correlations are 
quenched, so that protons and neutrons behave basically as rigid 
rotors. We have therefore  
\begin{eqnarray}
{\bar E}^{^(-)} \propto  \delta \omega_{0},\,\,\,\,\,\,\,\,\,
\Im_{sc}^{(-)} \simeq  \Im_{rig},
\end{eqnarray}
which yield
\begin{eqnarray}
m_{0}^{(-)} (M1) &= &\sum_{n_{-}} B_{n_{-}}^{(-)}  \propto  \Im_{rig} 
\delta\nonumber\\
m_{1}^{(-)} (M1)  &\simeq& {\bar E}^{(-)} 
\sum_{n_{-}} B_{n_{-}}^{(-)}  \propto  \Im_{rig} \delta^{2}. 
\label{mhighom}
\end{eqnarray}
According to the above formulas,
the superfluid to normal phase transition affects the deformation law.
While, in fact, the energy-weighted sum     
remains quadratic in the deformation,   
the behavior of the unweighted summed strength  with $\delta$
changes from quadratic to linear. 

We can therefore conclude that  the scissors M1 strength is 
closely correlated with the nuclear moment of inertia not only at low 
but also at large rotational frequency. 
More in detail, while the quasiparticle
energy moves downward because of the weakening of pairing, the $M1$ 
strength increases with $\Omega$ before backbending because of the  
increasing axial deformation and the gradual enhancement of the 
moment of inertia. Above the  backbending critical value, when 
the nucleus undergoes a transition 
from a superfluid to an almost rigid phase, as result of the 
the alignment of few quasiparticles with high spin,
the $M1$ strength jumps to a 
plateau, due to a sudden increase of the 
moment of inertia, while  
the deformation parameter $\delta$ 
remains practically constant.

Also the onset of triaxiality raises  
$m_{1}(M1)$  at large rotational frequency, to a modest extent. Indeed, from Eqs. 
(\ref{B12}) and (\ref{B3}) we get ($i=1,2,3$)
\begin{equation}
\sum_{i} E_{i}^{(-)} B_{i}^{(-)}(M1)\uparrow = \left(1 + \frac{5}{18} 
\sin^{2}{\gamma}\right)
m_{1}^{(-)} (M1)
\end{equation}
For $\gamma = 50^{0}$
the $m_{1}^{(-)} (M1)$ gets enhanced by a factor 1.16.
A further contribution comes from the changes in the shell 
structure induced by the rotation
which increase the number of configurations taking part  
to the motion over the whole energy range. The new configurations 
generate new transitions on one hand,  and, on the 
other hand, enhance the amplitudes of
collective as well non collective 
transitions.

\section{Conclusions}

Our analysis shows that fast rotation strengthens the coupling  
between  quadrupole  and monopole modes, broadens appreciably the isoscalar 
quadrupole giant resonance and washes out the isovector monopole and 
quadrupole peaks.  These rotation induced effects 
are found to be more appreciable than 
what predicted in Ref. \cite{Sh84}. On the other hand, the two 
approaches differ in several details. We accounted for the $\Delta N=2$ 
coupling in generating the Nilsson states and included the Galilean 
invariance restoring piece according to the prescription of Ref. 
\cite{Na96}. Moreover, we enforced the HB stability conditions, provided by 
Eq.(\ref{new}), that yield deformation parameters very close to the
self-consistent values and fixed the strength parameters of
the interaction so as to guarantee the separation of the spurious modes from the
intrinsic excitations at each rotational frequency.
 
The most meaningful and intriguing result of our calculation concerns 
the orbital, scissors-like, $M1$ excitations.
The enhancement of the overall $M1$ strength at high rotational 
frequencies emphasizes the dominant role of the scissors mode over 
spin excitations in fast rotating nuclei and represents an additional 
signature for superfluid to normal phase transitions in deformed 
nuclei. If confirmed experimentally, this feature would provide new 
information on the collective properties of deformed nuclei. 

\section*{Acknowledgments}
This work was partly supported by the Czech grant agency under the 
contract No. 202/02/0939, the Italian Ministero dell'Istruzione, 
Universit\'a and Ricerca (MIUR) and by the  
Grant No.\ BFM2002-03241 
from DGI (Spain). R. G. N. gratefully acknowledges support from the 
Ram\'on y Cajal program (Spain).


\begin{thebibliography}{99}
\bibitem{BoMoII} 
                 A. Bohr and B.R. Mottelson, {\it Nuclear Structure} 
                 (Benjamin, New York, 1975) Vol.II.	   
\bibitem{Dan58}
                M. Danos, Nucl.\ Phys.\ {\bf 5}, 23 (1958).
\bibitem{Oka58}
                K. Okamoto, Phys.\ Rev.\ {110}, 143 (1958).
\bibitem{Kish75}
                T. Kishimoto, J. M. Moss, D.H. Youngblood, J.D. 
                Bronson, C.M. Rozsa, D.R. Brown, and A.D. Bacher, 
		            Phys.\ Rev.\ Lett.\ {\bf 35}, 552 (1975).
\bibitem{oct92}
                T. Nakatsukasa, K. Matsuyanagi, and S. Mizutori,
                Prog.\ Theor.\ Phys\ {\bf 87}, 607 (1992).
\bibitem{n92}	
                R. Nazmitdinov and S. \AA berg,
	              Phys.\ Lett.\ B\ {\bf 289}, 238 (1992).

\bibitem{SuRo77}
                T. Suzuki and D.J. Rowe, 
		            Nucl.\ Phys.\ A\ {\bf 289}, 461 (1978).
           
\bibitem{ZaSpet78} 
                  D. Zawischa, J. Speth, and D. Pal, 
		              Nucl.\ Phys.\ A\ {\bf 311}, 445 (1978).
\bibitem{Ab87} 
                S. Aberg, 
		            Nucl.\ Phys.\ A\ {\bf 473}, 1 (1987).		  
\bibitem{Lo78} 
                N. Lo Iudice and F. Palumbo, 
	              Phys.\ Rev.\ Lett.\ {\bf 41}, 1532 (1978).
\bibitem{Boh84}
                D. Bohle, A. Richter, W. Steffen, A.E.L. Dieperink, 
	              N. Lo Iudice, F. Palumbo, and O. Scholten, 
	              Phys.\ Lett.\ B\ {\bf 137}, 27 (1984).
\bibitem{Lo00} 
                For an exhaustive list of reference N. Lo Iudice, 
                Rivista Nuovo Cimento {\bf 9}, 1 (2000).
\bibitem{RinSch} 
                P. Ring and P. Schuck, {\it The Nuclear Many-Body Problem}
                (Springer-Verlag, New York, 1980) for references.
\bibitem{KN}
               J. Kvasil and R.G. Nazmitdinov,
               Fiz.\ Elem.\ Chastits At.\ Yadra\ {\bf 17}, 613 (1986)
               [Sov.\ J.\ Part.\ Nucl.\ {\bf 17}, 265 (1986)].
\bibitem{Sh95}
               Y.R. Shimizu and M. Matsuzaki,
               Nucl.\ Phys.\ A\ {\bf 558}, 559 (1995).
\bibitem{Rob}
               L.M. Robledo, J.L. Egido, and P. Ring,
               Nucl.\ Phys.\ A\ {\bf 449}, 201 (1986).
\bibitem{Na87}
               R.G. Nazmitdinov,
               Yad.\ Fiz.\ {\bf 46}, 732 (1987)
               [ Sov.\ J.\ Nucl.\ Phys.\ {\bf 46}, 412 (1987)].
\bibitem{Na96} 
               T. Nakatsukasa, K. Matsuyanagi, S. Mizutori, 
	             and Y.R. Shimizu,
               Phys.\ Rev.\ C\ {\bf 53}, 2213 (1996).	
\bibitem{RMP}
               Y.R. Shimizu, J.D. Garrett, R.A. Broglia, M. Gallardo, 
               and E. Vigezzi, 
	             Rev.\ Mod.\ Phys.\ {\bf 61}, 131 (1989).
\bibitem{AF01} 
               D. Almehed D, F. D\"onau, S. Frauendorf, 
	             and R.G. Nazmitdinov,
               Phys.Scr. {\bf T88} 62 (2000);
               D. Almehed, S. Frauendorf, and F. D\"{o}nau,
               Phys.\ Rev.\ C\ {\bf 63} 044311 (2001). 	      

\bibitem{Sno86} 
                K. A. Snover, 
		            Ann.\ Rev.\ Nucl.\ Part.\ Sci.\ {\bf 36}, 
                545 (1986) and references therein.
\bibitem{Gaar}
                J.J. Gaardh\o je, 
                Ann.\ Rev.\ Nucl.\ Part.\ Sci.\ {\bf 42}, 
                483 (1992) and references therein. 
\bibitem{Sh84} 
                Y.R. Shimizu and K. Matsuyanagi, 
		            Progr.\ Theor.\ Phys.\ {\bf 72}, 1017 (1984);
		            {\it ibid} {\bf 75}, 1167 (1986).
\bibitem{KvNa2} 
                see for instance J. Kvasil and R.G. Nazmitdinov, 
	              preprint nucl-th/0402031, to be published
		            in Phys.\ Rev.\ C\ ;
                J. Kvasil, R.G. Nazmitdinov, and A.S. Sitdikov,  
                to be published in Yadernaya Fizika .
                	
\bibitem{DeHe89} C. De Coster and K. Heyde, Phys. Rev. Lett.  
                {\bf 63}, 2797 (1989).
                
\bibitem{DeHe91} C. De Coster and K. Heyde, Nucl. Phys.   
                {\bf A 524}, 441 (1991).	
\bibitem{Kv98} 
                J. Kvasil, N. Lo Iudice, V.O. Nesterenko, and M. Kopal,
                Phys.\ Rev.\ C\ {\bf 58}, 209 (1998).
		
\bibitem{SaKi89}
                H. Sakamoto and T. Kishimoto,
                Nucl.\ Phys.\ A\ {\bf 501}, 205 (1989).
       
\bibitem{Lo96} N. Lo Iudice,
               Nucl.Phys. {\bf A 605}, 61 (1996).
\bibitem{A03}
               D. Almehed, F. D\"onau, and  R.G. Nazmitdinov,
               J.\ Phys.\ G:\ Nucl.\ Part.\ Phys.\ {\bf 29}, 2193 (2003).

\bibitem{Na02}
               R.G. Nazmitdinov, D. Almehed, and F. D\"onau,
               Phys.\ Rev.\ C\ {\bf 65}, 041307(R) (2002).

\bibitem{Ma76}
               E.R. Marshalek,
               Nucl.\ Phys.\ A\ {\bf 266}, 317 (1976).
\bibitem{Ja90} 
                A.K. Jain, R.K. Sheline, P.C. Sood, and K. Jain, 
                Rev.\ Mod.\ Phys.\ {\bf 62}, 393 (1990).
\bibitem{Wyss}
               R. Wyss, W. Satula, W. Nazarewicz, and A. Johnson,
               Nucl.\ Phys.\ A\ {\bf 511}, 324 (1990).    
\bibitem{Sa99} 
               R.Ch. Safarov and A.S. Sitdikov, Izv. A.N. {\bf 63},                 
		           162 (1999) and references therein.	 
\bibitem{Frau}
               S. Frauendorf and F.R. May,
               Phys.\ Lett.\ B\ {\bf 125}, 245 (1983).
\bibitem{Ca76} 
                B. Castel and I. Hamamoto, 
                Phys.\ Lett\ B\ {\bf 65}, 27 (1976).
\bibitem{LoRi89} 
                N. Lo Iudice and A. Richter
                Phys.\ Lett.\ B\ {\bf 228}, 291 (1989).
\bibitem{Rich93} 
                A. Richter,
                Nucl.\ Phys.\ A\ {\bf 553}, 417c (1993) 

\bibitem{Pa85} 
                F. Palumbo and A. Richter, 
	              Phys.\ Lett.\ B\ {\bf 158}, 101 (1985).                
\bibitem{Lo85} 
                N. Lo Iudice, E. Lipparini, S. Stringari, 
                F. Palumbo, and A. Richter, 
                Phys.\ Lett.\ B\ {\bf 161}, 18 (1985).
\bibitem{ZaZhe91} 
                L. Zamick and D.C. Zheng, 
		            Phys.\ Rev.\ C\ {\bf 44}, 2522 (1991).	       	       
\bibitem{Lo98}  
                N. Lo Iudice,
                Phys.\ Rev.\ C\ {\bf 57}, 1246 (1998).
\bibitem{LoRi93} 
                N. Lo Iudice and A. Richter
                Phys.\ Lett.\ B\ {\bf 304}, 193 (1993).
\bibitem{Piet98} 
                N. Pietralla, P. von Brentano, R.-D. Herzberg, U. Kneissl,
                N. Lo Iudice, H. Maser, H. H. Pitz, and A. Zilges,  
                Phys.\ Rev.\ C\ {\bf 58}, 184 (1998).
\bibitem{Zi90} 
               W. Ziegler, C. Rangacharyulu, A. Richter, and C. Spieler,
               Phys.\ Rev.\ Lett.\ {\bf 65}, 2515 (1990).
\bibitem{Ma93} 
               J. Margraf, R.D. Heil, U. Kneissl, U. Meier, H.H. Pitz,
               H. Friedrichs, S. Lindenstruth, B. Schlitt, C. Wesselborg, 
               P. von Brentano, R.-D. Herzberg, and A. Zilges,
               Phys.\ Rev.\ C\ {\bf 47}, 1474 (1993).	       
\end{thebibliography}
\end{document}